# Fabrication Optimization of van der Waals Metasurfaces: Inverse Patterning Boosts Resonance Quality Factor

*Jonas Biechteler[1], Connor Heimig[1], Thomas Weber[1], Dmytro Gryb[1], Luca Sortino[1], Stefan A. Maier[2, 3], Leonardo de S. Menezes[1,4], Andreas Tittl[1*]*

[1]Chair in Hybrid Nanosystems, Nano-Institute Munich, Department of Physics, LMU Munich, Germany
[2]School of Physics and Astronomy, Monash University, Clayton, Victoria 3800, Australia
[3]Department of Physics, Imperial College London, London SW7 2AZ, United Kingdom
[4]Departamento de Física Universidade Federal de Pernambuco, 0670-901 Recife-PE, Brazil
*Corresponding author. Email: Andreas.Tittl@physik-uni.muenchen.de

**Abstract**

Van der Waals (vdW) materials have garnered growing interest for use as nanophotonic building blocks that offer precise control over light-matter interaction at the nanoscale, such as optical metasurfaces hosting sharp quasi-bound states in the continuum resonances. However, traditional fabrication strategies often rely on lift-off processes, which inherently introduce imperfections in resonator shape and size distribution, ultimately limiting the resonance performance. Here, an optimized fabrication approach for vdW-metasurfaces is presented that implements inverse patterning of the etching mask, resulting in increased resonator quality solely limited by the resolution of the electron beam lithography resist and etching. Applying this inverse fabrication technique on hexagonal boron nitride (hBN), quality (Q) factors exceeding $10^3$ in the visible spectral range were demonstrated, significantly surpassing previous results shown by lift-off fabricated structures. Additionally, the platforms potential as a biosensor was displayed, achieving competitive sensitivity and figure of merit of 220 in a refractive index sensing experiment. The inverse technique was applied to create chiral metasurfaces from hBN, using a two-height resonator geometry to achieve up to 50 % transmittance selectivity. This inverse lithography technique paves the way towards high-performances vdW-devices with high-Q resonances, establishing hBN as a cornerstone for next-generation nanophotonic and optoelectronic devices.

# 1. Introduction

The unique optical properties of two-dimensional (2D) materials have drawn significant attention in nanophotonics ever since the discovery of graphene[1]. They are crystalline solids that consist of individual covalently bonded, atomically thin sheets that are held together in the stacking direction merely by van der Waals (vdW) forces. This structure enables the isolation of single flakes with arbitrary thickness ranging from bulk form down to the single atom level by simple mechanical exfoliation directly from the crystal. 2D or more broadly vdW materials offer rich physics, including Moiré photonics[2], room temperature single-photon emission[3,4], 2D magnetism[5] and spintronics[6].

The comparatively weak vdW forces between layers enable sophisticated stacking of different 2D-materials without lattice mismatch, creating possibilities for combinations of many individual materials and their properties.[7] This led to a special interest in hexagonal boron nitride (hBN) as an ideal substrate[8]. Its wide, indirect bandgap of about 6 eV and atomically smooth surface enable the efficient encapsulation of graphene, transition metal dichalcogenides (TMDC) or other optically active 2D materials.[9,10] Moreover, it reduces the impact of charge impurities and ambient exposure on the encapsulated material[11], thereby boosting the electrical (mobility) and optical properties[12]. Recently, hBN has also become the center of attention as a nanophotonic building block.[13–15] Its nearly lossless nature in bulk form from the ultraviolet to the near-infrared (NIR) spectrum favors the creation of optically resonant systems like waveguides or metasurfaces (Figure 1a).

Simultaneously harnessing hBN as an excellent encapsulation material and an optical resonator medium has tremendous potential.[16] By patterning hBN heterostructures into optically resonant devices, we can effectively shape the electromagnetic fields around the encapsulated medium within.[17] Furthermore, significant attention within 2D materials research has been given to the phenomenon of chirality[18,19], as it plays a crucial role in the investigation of excitonic valleys in TMDC monolayers. Merging hBN as an ideal substrate for these materials with the physics of ultra-sharp chiral qBICs holds the potential to enable experimental observation of intriguing phenomena, such valley polariton-condensation and subsequent valley lasing[20,21]. Therefore, development of high quality hBN-based platforms is required to extend the tools available for further research in this area.

Optical metasurfaces offer precise control over light manipulation through periodically arranged sub-wavelength building blocks, achieving effects beyond conventional optics.[22] They can host symmetry-broken quasi-bound states in the continuum (qBICs), which have been shown to offer great control over light matter coupling[23]. By tuning asymmetry within each meta-atom, the radiative losses of the system can be directly tailored, resulting in arbitrarily sharp resonances and enabling strong light confinement within each resonator (Figure 1b).

Previous nanofabrication approaches with vdW materials have typically relied on electron-beam lithography (EBL) followed by the deposition of a metal hard mask and a wet-chemical lift-off[24–28]. However, especially on the atomically flat and dangling bond free surfaces of vdW materials, lift-off can create inevitable imperfections in the metal mask. These appear in form of deformations and deviations between mask parts and translate to the resonator shape during reactive-ion etching of the underlying medium, unintentionally increasing intrinsic losses in the final structures, hence making this process inherently faulty (Figure 1c).[29] Because of these fabrication issues, previous vdW-based metasurfaces have struggled to achieve hihg quality factor (Q factor, defined as the resonance wavelength divided by the linewidth) resonances. Our goal is to eliminate these fabrication-induced limitations in order to maximize resonance signal quality and maximizing the system's performance.

In this work, we demonstrate an inverse lithography fabrication approach optimized for the nanostructuring of hBN and other vdW materials (Figure 1d), with accuracy limited solely by the e-beam resist resolution and the subsequent fully dry etching steps. We find a significantly decreased variation in the resonator size distribution, quantified by a 65 % reduction in standard deviation compared to lift-off fabricated metasurfaces. This reduction leads to enhanced optical performance with quality factors exceeding $10^3$ in the visible and up to 7-fold increase compared to previous hBN metasurface works[24,25], while preserving high modulation depths of the resonances. These results represent a crucial advancement for integrated visible-range photonics, further extending hBN as an exciting photonic material platform. Furthermore, we demonstrate the versatility of our metasurfaces with a proof-of-concept refractometric biosensing experiment (Figure 1e), achieving competitive figures of merit of 220 driven by the high quality factors of the metasurfaces. Finally, we harness our fabrication technique in a two-step process to create advanced multi-height resonator designs for maximizing chirality (Figure 1f). We achieve formation of a pronounced chiral qBIC mode, displaying the potential of our hBN metasurfaces for valley-related applications or investigations in heterostructure geometries.

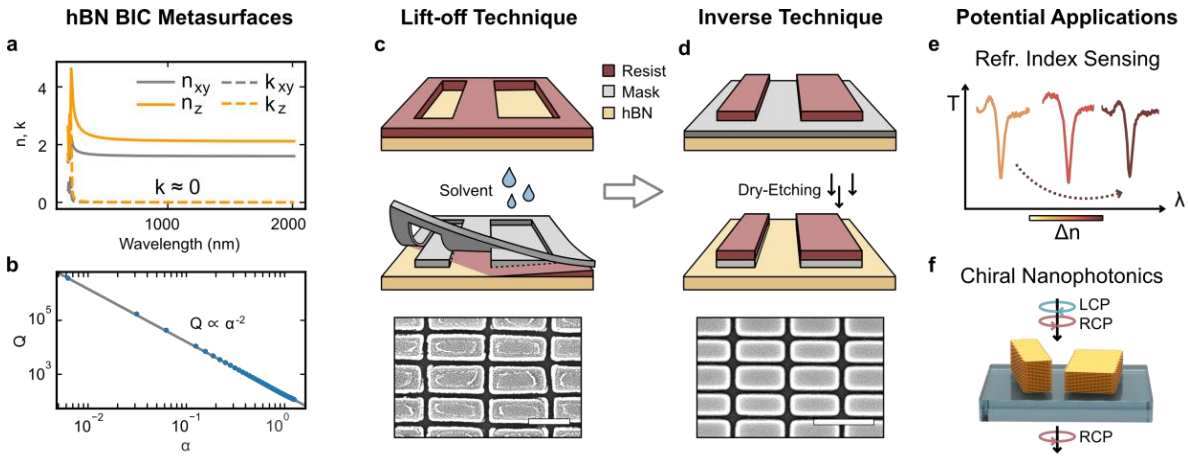

**Figure 1. Unlocking new possibilities for hBN metasurfaces using inverse patterning. a.** In-plane (xy) and out-of-plane (z) components of the refractive index n and extinction coefficient k. Note that k is approximately zero over the entire visible and Near-IR spectrum.[30] **b.** Q factors extracted from simulated spectra of qBIC metasurfaces from hBN with decreasing asymmetry $\alpha$. Q factors show inverse proportionality to α squared. **c.** Illustration of a lift-off process to form a metal hard mask for reactive-ion etching (top). SEM picture of a hBN metasurface fabricated by lift-off (bottom). **d.** Illustration of the inverse approach forming the metal hard mask by etching (top). Associated SEM picture (bottom) shows clear improvement in homogeneity and size continuity. **e.** Showcase of a refractive index sensing measurement. Shifting resonance spectra indicate a change in metasurface surrounding refractive index. **f.** Illustration of a single unit cell of a maximally chiral metasurface.

## 2. Results

**Fabrication Approach Comparison**

Our optimized fabrication technique deviates from previous lift-off-based techniques in how we approach the creation of a metal hard mask for subsequent etching. Common lift-off strategies rely on deposition of the mask material film on top of a positive EBL resist (such as PMMA or CSAR), which has been patterned to form holes where the mask material is intended to directly coat the active material. Following this, a solvent is used to remove the unexposed resist areas underneath the mask film and selectively peeling off the film parts without direct contact to hBN (for details see Supporting Information Note 1). This can lead to a series of problems that can be avoided by bypassing the lift-off process in its entirety.

In our inverse technique, we first coat the hBN flakes with a layer of chromium (Cr) followed by an EBL resist (CSAR 62) (Figure 2a, step I). In general, positive resist breaks down in areas exposed to the electron-beam, leading to the creation of holes in these areas during the subsequent wet-chemical development. Therefore, in lift-off approaches, one would normally expose the resist in areas to form holes at the intended locations and in the shape of the resonators. Instead, during EBL (step II), we expose the gap areas, exactly inverse to what is usually written, so after development a pattern of the desired gaps is being formed. We avoid the usage of negative resists here, as they either require harsh, wet-chemical removal, or do not offer on par resolution with positive resists[31,32], which can be easily removed by oxygen-plasma exposure and provide sufficient etch-stability for the following step. The resist layer now acts as etching mask (step III) for the hard mask film. As a mask material, we chose materials like Cr in this case or alternatively silicon dioxide ($SiO_2$) that can be selectively etched with a plasma based on chlorine ($Cl_2$) or sulfur hexafluoride ($SF_6$), without aggressively removing resist or hBN. These materials also offer the required sturdiness to withstand the following hBN etching (step IV) and can be easily removed by a final $Cl_2$ plasma etching step (step V) to finalize the hBN metasurfaces.

Our inverse technique offers several practical advantages, avoiding issues unique to lift-off that can ultimately compromise the entire fabrication process. First, liquid exposure during lift-off can potentially shift, rotate or remove resonator mask parts that poorly adhere to the atomically flat vdW material's surface. Furthermore, lift-off requires a highly directional deposition of the mask material on the patterned resist layer. Coating with slight deviations from normal angle can leads to deposition of material on the resist sidewalls, connecting the different mask parts. This leads to uncontrollable breaking mechanics at the mask outline that induce drastic inhomogeneity and translate to the resonator shape. Coated sidewalls can also block solvent from fully dissolving resist areas, leading to an incomplete lift-off. With our technique, the hard mask can be directly deposited onto the 2D material flakes, enabling further deposition methods such as sputtering. This creates a uniform, high-quality layer that pins down the 2D material flakes, and can be patterned with a single resist layer on top.

Our metasurface design employs a double rod BIC unit cell. By increasing the width of one rod and reducing the width of the other by a given value $\Delta w$, we can introduce an asymmetry parameter $\alpha = \Delta w/w_0$ (Figure 2b). Importantly, this choice keeps the total resonator volume constant with changing asymmetry. After optimization in numerical simulations, we chose to set the gaps between all hBN rods to 60nm and select a flake height around 150nm. This choice provides sufficient resonator volumeto ensure spectral separation between the qBIC resonance and the grating mode, while still avoiding the emergence of higher order qBIC resonances (see Supporting Information Note 2).

To compare common lift-off techniques with our inverse technique, we employed both fabrication approaches to produce a set of metasurfaces from single hBN flakes. For both cases, a series of scanning-electron microscopy (SEM) images was taken to assess resonator quality and size distribution. The SEM images were analyzed using a contour matching algorithm (Figure 2c).[33] Specifically, for each set of SEM images from one metasurface we evaluate the distributions of widths of both the wider ($w_1$) and narrower rods ($w_2$) and calculate the means ($\mu_{w1}$, $\mu_{w2}$). Next, we combine each distribution by shifting their means individually to zero and merge them into one dataset. This gives us a statistical insight into how the two fabrication methods produce geometrical deviations from resonator to resonator. For the lift-off technique, we find a large spread of the geometrical parameters around the mean with a standard deviation of $\sigma_{LO} = 5.42$ nm (Figure 2d). In contrast, our inverse approach yields a 65 % decreased standard deviation of $\sigma_{Inv} = 1.91$ nm, an significant boost in fabrication accuracy. These improvements not only reduce intrinsic losses[33], but also enable the patterning of smaller asymmetries in qBIC metasurfaces – and thus generating resonances with higher Q factors - that would otherwise be obscured by statistical variations in resonator size.

We note that the standard deviation value of our lift-off fabricated hBN-based metasurfaces is larger than a previous analysis on silicon-based metasurfaces[33]. We attribute this deviation to the additional challenges associated with nanofabrication of 2D materials, such as the atomically flat surface of hBN, which provides less adhesion than silicon surfaces and increases chances of slight rotations and movement of hard mask material during lift-off. In contrast, our inverse technique does not require good adhesion of the hard mask, since it only requires withstanding a series of liquid-free reactive-ion etching steps.

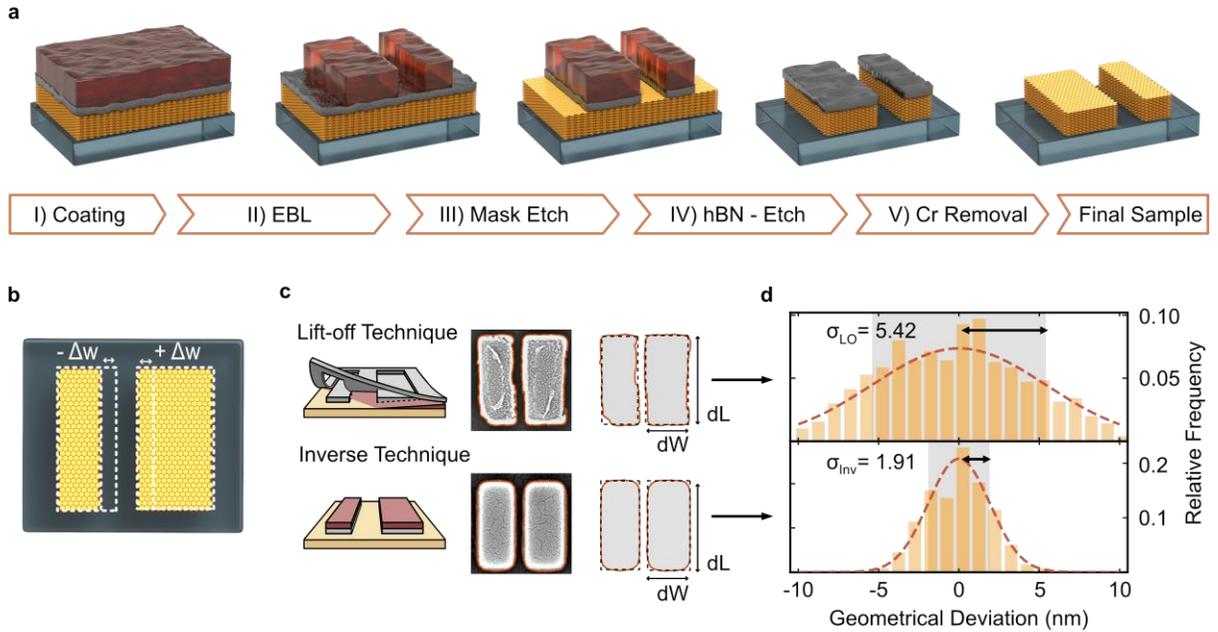

**Figure 2. Evaluation of the inverse writing technique. a.** Illustration of the fabrication process flow in the inverse technique. After coating of the hBN flake with chromium and EBL resist (I), the later gaps are exposed via EBL and washed away during development. This leaves unexposed

resist with resonator dimensions (II) which can now be used as a mask for reactive ion etching of the Cr film (III). After subsequent resist removal via oxygen plasma, hBN can be etched (IV) and the Cr removed by dry etching (V). **b.** Illustration of the unit cell design of choice. The double rod structure induces asymmetry by subtraction and addition in width $\Delta w$, leaving the total resonator volume unchanged. **c.** Showcase of the SEM image fitting method employed for statistical analysis of resonator geometry for both inverse and lift-off technique. **d.** Distribution of resonator geometry deviation for the two approaches, both fitted by a Gaussian curve.

**Optical Measurements of hBN Metasurfaces**

To experimentally validate the effects of our optimized fabrication technique, we create a set of metasurfaces with decreasing values of α ranging from $\alpha = 27.5\ \%$ down to $\alpha = 3.5\ \%$ and perform measurements in an optical spectroscopy setup (for details see Methods). The collected transmittance data (Figure 3a) shows qBIC resonances with increasing sharpness and a slight redshift for decreasing $\alpha$. The Q factors (Fig 3a, inset) were extracted using a temporal coupled mode theory approach[34] and reach up to 1350. We find the lowest achievable asymmetry to be 5 nm or $\alpha = 3.5\ \%$ of the resonator width, before the resonance becomes no longer detectable. qBIC resonances of metasurfaces with lower α are not resolvable, as α becomes too similar in size to the statistical geometric deviation for each resonator that we extracted earlier.

Furthermore, we fabricate a set of metasurfaces with $\alpha = 5\ \%$ and increasing in-plane scaling-factor *S* to cover the entire visible spectrum (Figure 3b). Here, we measure a series of pronounced, strongly modulated resonances and extracted Q-factors in the order of $10^3$, peaking at 2085, a 7-fold increase to previous results from hBN qBIC metasurfaces[24,25].

Over the full 450 nm to 900 nm range, our metasurfaces maintain excellent Q factor stability (Figure 3b, bottom), thereby outperforming previous metasurfaces by around a factor of 5 on average. Furthermore, we achieve this performance without a notable decrease in modulation, which is especially noticeable at the edges of the range[24]. Previous approaches have struggled to maintain high Q factors at both the ultraviolet and NIR ends of the spectrum. Since smaller structures are a necessity towards the lower end of the spectrum, the edge quality and size variations of resonators become a crucial factor that we directly tackle with the advantages of our inverse technique. At the other end of the range, pushing the resonance to longer wavelengths requires larger resonator volume to reach the required optical mode volume. In the past, the height of the etching mask was limited by lift-off approaches, ultimately restricting the height of the hBN flake that could be etched. In contrast, the inverse technique enables etching and creation of masks with significantly increased height suitable for etching thicker flakes, enabling further improvements of NIR resonance performance.

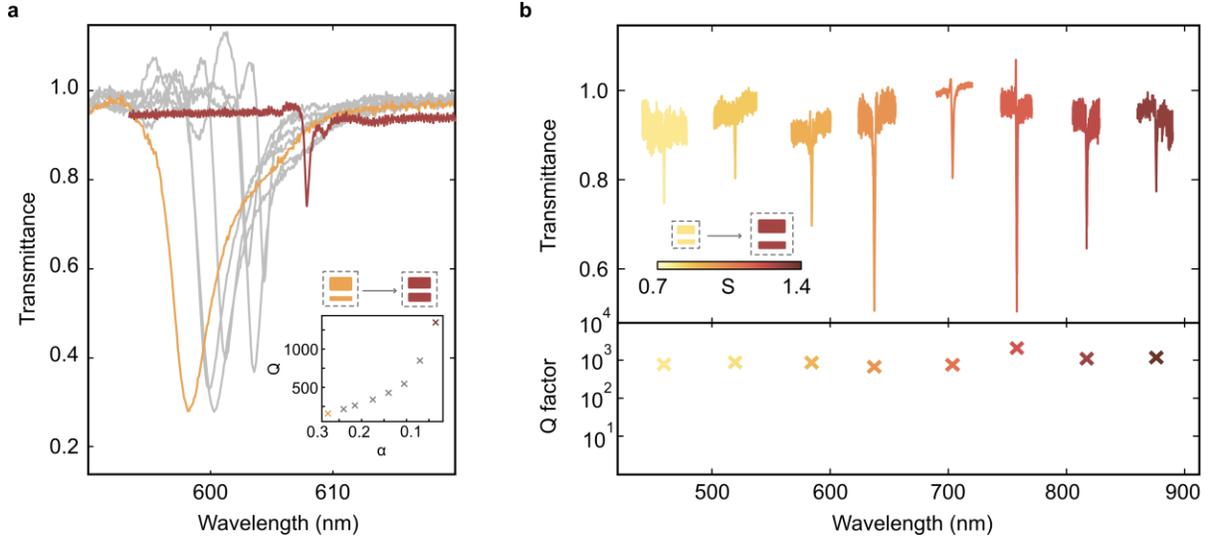

**Figure 3. Optical measurement of hBN metasurfaces. a.** Transmittance spectra of metasurfaces with decreasing asymmetry with extracted Q factors (inlet). **b.** Spectra of qBIC resonances of different metasurfaces with increasing 2D-Scalingfactor (top) and fitted Q factors (bottom). $S = 1.0$ represents an in-plane period of $p_x = p_y = 400$ nm with resonator gaps to all sides of 60 nm.

**Bulk Refractive Index Sensing**

The pronounced and sharp resonances produced by our hBN metasurfaces fabricated by inverse technique have great potential for potential ultra-sensitive bio detection systems[35,36]. Here we focus on characterizing the sensing properties of our metasurfaces and evaluate their bulk refractive index sensing capabilities.

We performed a series of measurements on an inverse fabricated metasurface exhibiting high Q factor (>1300) and large resonance amplitude. We then immersed it in water-glycerol mixtures with different concentrations and find a clear shift in resonance position $\Delta\lambda$ with changing concentrations of glycerol in water as well as a corresponding change in metasurface surrounding refractive index $\Delta n_{env}$ (Fig 4a). We extract the shifted peak positions and plot them in relation to the respective $\Delta n_{env}$ to find the associate Bulk Sensitivity $S_B = \Delta\lambda/\Delta n_{env} = 118$ nm/RIU (Figure 4b).

To determine the limit of detection of our sensor, we first evaluate the variation of the resonance position over time, which corresponds to the noise of our system (see Supporting Information Note 4). Based on the measured noise value of $\sigma = 2.258 * 10^{-3}$ nm, we calculate a limit of detection (LOD) for our sensor[37,38] of LOD $= 3\sigma * S_B^{-1} = 5.74 * 10^{-5}$ RIU. This ultra-fine sensing limit, compared to our smallest refractive index variation $\Delta n_{env} = 3.5 * 10^{-3}$ RIU (Figure 4c) in experiment, could warrant further investigations in more sophisticated experiments using a microfluidic channel. Furthermore, the central advantage of our metasurfaces is the high Q factor and corresponding small average values of the full width at half maximum $FWHM = 0.535$ nm (Figure 4d) during the experiment, giving rise to a figure of merit FOM $= S_B/FWHM = 220$.

It is important to note that on the one hand plasmonic sensors based on metallic nanostructures are often capable of providing higher bulk sensitivities $S_B > 1000$ nm/RIU [39,40], enabled by the strong electromagnetic fields localization at metal/dielectric interfaces. On the other hand, the unavoidable intrinsic losses associated with established plasmonic sensors keep possible Q factors down and restrict achievable FOMs and subsequently LODs. Dielectric metasurfaces driven by qBICs benefit from maximizing the Q-factor by applying structural changes like shallow etching, that minimize intrinsic losses and achieve extremely sharp resonances[41]. In experiment, these high Q-factors enable higher FOMs and LODs (FOM > 800, LOD = $3.01 * 10^{-5}$ RIU), but simultaneously, the applied geometric changes restrict the sensing volume and concurrently lower achievable sensitivity ($S_B = 26$ nm/RIU) compared to our work. Conversely, other studies with dielectrics focused on enhancing near-field sensitivity ($S_B = 407$ nm/RIU) deliver only much lower FOMs (FOM = 26) due to poor Q factors [42].

The ability of our sensor to simultaneously achieve high sensitivity and a high FOM positions it as a competitive alternative to current refractometric metasurface sensors in the field. Additionally, other demonstrations with extremely high Q factors have mostly been presented at NIR wavelengths or in simulations. Reaching competitive performance in the visible is particularly challenging due to the required fabrication accuracy, which our inverse technique successfully delivers.

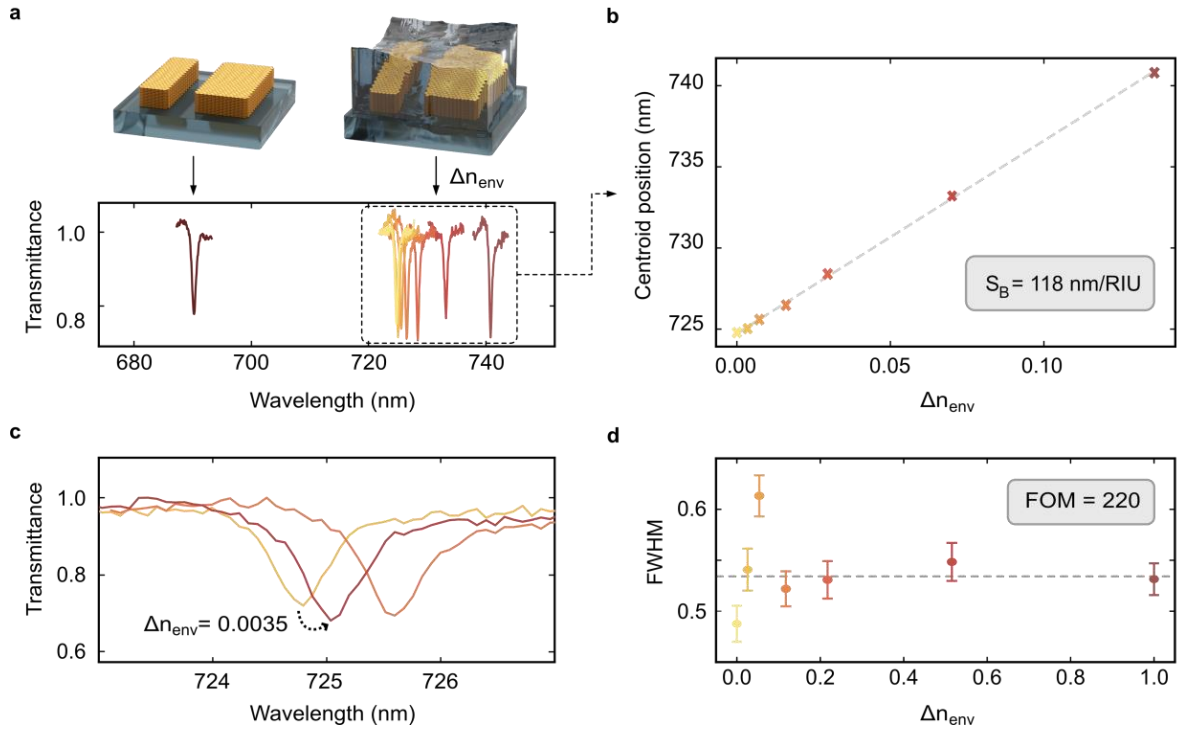

**Figure 4. Refractive index sensing. a.** Illustration of the transmittance measurements in air (left) and liquid (right) with measured transmittance spectra (bottom). **b.** Extracted centroid positions of qBIC spectra measured under different glycerol-water mixtures. Measurement gives a sensitivity $S_B = 117$ nm/RIU. **c.** Shift of resonance spectra by small change in refractive index surrounding. **d.** Full-width-half-maximum extracted from all spectra from the sensing measurements, with an average of FWHM = 0.535 nm resulting in an average FOM = 220.

**Maximally Chiral Metasurfaces**

Advancing nanoscale light control through polarization selective light–matter interactions is essential for applications in chiral quantum optics.[43] In this context, we aim to extend the functionality of hBN as a chiral metasurface cavity.[44] Previous studies struggled to achieve pronounced resonances from planar structures that are required to study strong interactions with chiral near fields.[25,45] Adding an out-of-plane component to metasurfaces has been shown to be a promising solution for this problem, strongly enhancing the degree of chirality of qBIC resonances and attaining improved circular dichroism[46]. Yet, this approach requires a multi-step EBL process that leaves little room for shifts in alignment between individual steps. Our fabrication technique is well suited for the use in multi-step EBL processes, as it allows for seamless nanostructuring of uneven or pre-patterned materials, given that mask coating on top creates a more uniform surface to structure EBL resist on. This gives the freedom to reorder the steps and lets us first apply the out-of-plane components before patterning all resonators in a single step, eliminating chances of misalignment.

As a first step (Figure 5a), we use our technique to structure the hBN flake's surface with a low contrast grating that will give later resonators the desired height difference of about $60\ nm$. After etching and removal of the first mask, a second mask film is deposited. We then use our inverse technique to pattern the desired resonator shapes, spatially aligned to our previously written low contrast grating. Note, that this sequence of actions improves EBL resist coating, compared to writing the resonators first and applying the height difference afterwards. It also allows us to pattern all resonators in a single EBL step, providing perfect unit cell and resonator alignment. The final unit cell design consists of a double rod geometry with a given height difference, where both rods are tilted to have an opening angle theta respect to each other. We tune both parameters to create and optimize maximally chiral qBICs in our metasurface (see Supporting Information Note 5). Via numerical simulations, we tailor these parameters to minimize light-matter coupling between RCP light and our structures (Figure 5b), while LCP light of resonant wavelength interacts optimally and creates the desired maximally chiral near fields. The fabricated chiral metasurfaces again show great fabrication accuracy and reproduction of the target design parameters (Figure 5c).

We characterized the metasurface by measuring the co-polarized ($T_{RR}$, $T_{LL}$) and cross-polarized ($T_{RL}$, $T_{LR}$) transmittance spectra (Figure 5d). Here, $T_{ij}$ means the amount of i-circularly polarized light transmitted as j-circularly polarized light. On the one hand, we observe the formation of a strong resonance in $T_{LL}$, with over 50 % modulation. On the other hand, $T_{RR}$ shows almost no resonance, demonstrating prominent, handedness-selective qBIC mode formation. For the cross-polarized terms, we measure negligible signals, showing that the system fully suppresses unintended polarization scrambling, indicating strong potential for applications as a sharp polarization maintaining filter.

We calculate a maximum linear circular dichroism (CD) by $\text{CD} = (T_{LL} - T_{RR})/(T_{LL} + T_{RR}) = 0.375$ at resonance spectral position, which is over 30 times larger than the values reported in comparable works on chiral hBN systems relying on plane metasurfaces[25]. Our results pave the way towards application as a maximally chiral cavity to boost chiral light-matter coupling in a vdW-heterostructure or as a chiral filter.

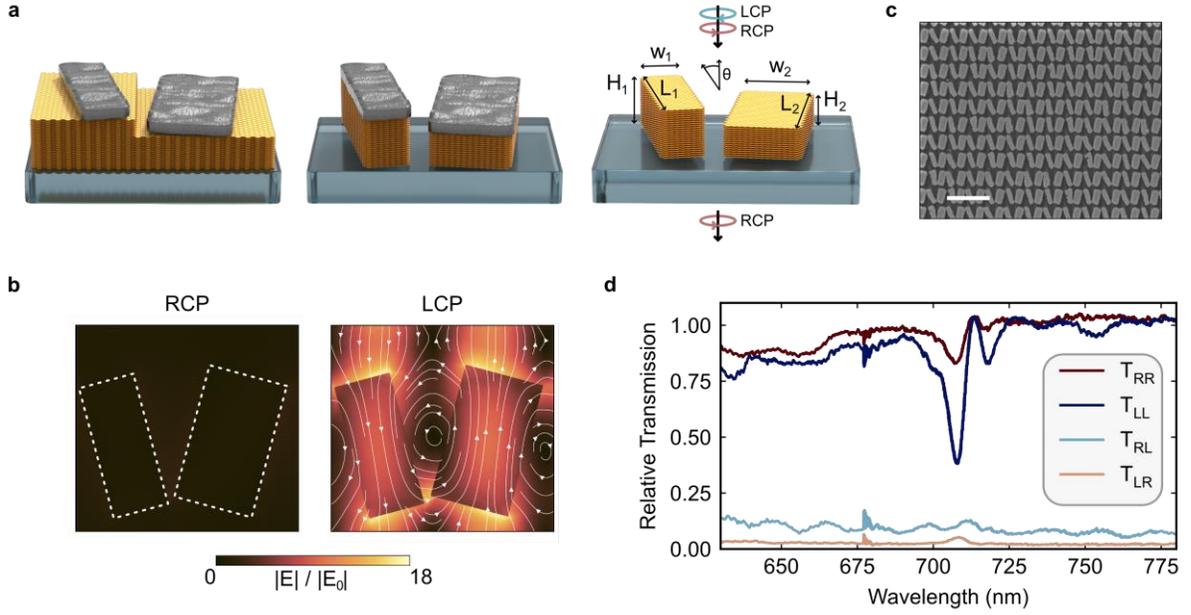

**Figure 5. Realization of chiral hBN metasurfaces. a.** Illustration of the extra fabrication steps to achieve multi-height asymmetry. hBN flake is firstly patterned to two heights in every unit cell. Secondly, resonator shape is patterned into chromium in one step by inverse technique (left) and subsequently etched using reactive-ion etching (middle). Illustration of the geometry of the final two-height resonator unit cell (right). The unit cell design has a period of 500 nm, rods tilted by $\theta = 15°$, with lengths of $L_1 = 380$ nm and $L_2 = 345$ nm, widths $W_1 = 125$ nm and $W_2 = 180$ nm, and heights $H_1 = 220$ nm and $H_2 = 160$ nm, respectively. **b.** Simulation of the electric fields in the unit cell excited by RCP (left) and LCP (right) light. **c.** SEM picture of a chiral hBN metasurface. **d.** Co- and cross-polarized transmittance spectra. Co-polarized spectra show the emergence of a strong qBIC resonance for only one handedness, cross-polarized terms are insignificant.

In conclusion, we have presented an optimized fabrication approach for vdW-based nanophotonic systems that relies on direct coating and nanostructuring of an etching mask, circumventing all problems associated with a common lift-off. It can be directly applied for any substrate or surface to nanostructure and gives immediate improvement in process reliability and quality of outcome. Utilizing a mask layer directly on top of 2D materials like hBN and patterning it first to create an etching mask, not only pins down the flake but also circumvents the need for strong surface adhesion, that naturally cannot be provided on a dangling bond free, atomically smooth surface. Our approach decreases the standard deviation of resonator dimension distribution compared to structures made with common lift-off technique by up to 65 %, effectively limiting the attainable fabrication resolution to only the EBL resist. This results in direct improvement of metasurface performance in all operational spectral regions, while also enabling the detection of qBIC signal from finer asymmetries and realizing Q-factors up to >2000. This ultimately expands the scope of application of our hBN platform to novel territories.

The proof-of-concept refractive index sensing experiments show competitive scores in bulk sensitivity and particularly FOM, empowered by the fine linewidths of our qBIC resonances which enhance the precision and effectiveness of the sensor. The pronounced contrast in signal we measure further underscores the potential of hBN, giving prospects as a possible biodetection system in the visible spectrum.

Additionally, our fabrication approach can seamlessly be integrated in multistep EBL processes, simplifying multi-height patterning. This sets up successful realization of maximally chiral qBIC resonances in hBN, demonstrating a robust method for controlling light chirality at the nanoscale. These advancements open exciting opportunities for integrating hBN-based chiral metasurfaces into van der Waals heterostructures, where the pronounced chiral resonances could enhance light-matter interactions for next-generation quantum optics, valley photonics and chiral sensors.

**Experimental Methods**

*Simulations*

Simulations of hBN metasurfaces were performed using the software CST Studio Suite 2021. For the $SiO_2$ substrate the refractive index was set to 1.45 and for hBN refractive index literature values were used[24].

*Fabrication*

Fused silica substrates were cleaned by sonication for 3 minutes in first acetone and then isopropanol and finally oxygen-plasma treated for 10 minutes. hBN was mechanically exfoliated from commercial bulk crystals (HQ Graphene) onto the substrate at a temperature of 160 °$C$ to evaporate moisture and promote larger flake release from weak, melted tape glue. The exfoliated flakes where then again cleaned in an oxygen plasma by 10 minutes to eliminate glue residues. Suitable flakes were selected by optical microscopy and their exact height measured with a profilometer (Brucker Dektak XT).

The flakes and substrate were then coated by 40 nm of chromium (alternatively 80 nm $SiO_2$) to create the mask film. On the Cr-coated sample an adhesion promoter (Surpass 4000) was applied before spin coating the EBL resist (CSAR 62, further diluted with anisole to a 1:1 ratio) at 2000 rpm to create an approximately 200 nm thick layer.

The resist was inversely exposed (E-beam on gaps) using electron beam lithography (Raith eLine plus) at 20 kV, aperture size 15 µm and developed in amyl acetate for 60 seconds.

These EBL parameters were used for inverse and lift-off fabricated samples.

The etching was operated in the with the following recipes:

- Cr-etching: 42 sccm $Cl_2$, 10 sccm $O_2$ at 12 mTorr. 20 W HF, 500 W ICP
- hBN-etch: 10 sccm $SF_6$, 5 sccm Ar at 6 mTorr. 300 W HF, 150 W ICP

First, the Cr layer was etched to create the etching mask for the hBN process that was subsequently performed. Remaining EBL resist was then removed by Oxygen Plasma treatment. The Cr mask was removed by Cr-etching to finish the sample.

*Optical characterization*

Linear transmittance spectra were acquired using a commercially available white-light transmission microscopy (Witec Alpha series 300). The substrates and metasurfaces were illuminated by a linearly polarized white-light source (Thorlabs OSL2) and light was collected with a 50x, numerical aperture $NA = 0.8$ objective. The signal was focused into a multimode fiber and sent in a grating-based spectrometer (density 600 grooves/mm, Si-CCD sensor). Spectra were referenced to the $SiO_2$ substrate background. In-liquid measurements were likewise performed using a 63x, NA = 0.8 immersion objective.

For chiral transmittance measurements, a fiber-coupled supercontinuum white light laser (SuperK FIANIUM from NKT Photonics) was employed and first directed through a polarizing beam splitter. A quarter wave plate (RAC4.4.20, B-Halle) was used to generate right- or left-circular polarized light from the horizontal or vertical linearly polarized components and the signal was subsequently collected using a 60x, $NA = 0.7$ objective. Co- and cross-polarized terms were measured using a chiral analyzer, consisting of a quarter wave plate (AQWP05-580, Thorlabs) plus a linear polarizer (WP25M-UB, Thorlabs). The chiral analyzer was placed after the collection objective and light directed to a spectrometer (Princeton Instruments) using a multimode fiber (Thorlabs M15L05, core size: 105 μm, $NA = 0.22$)

**Data availability**

The data that support the findings of this study are available from the corresponding author upon reasonable request.

**Supporting Information**

Supporting Information is available from the Wiley Online Library or from the author.

**Acknowledgments**

Funded by the European Union (ERC, METANEXT, 101078018). Views and opinions expressed are however those of the author(s) only and do not necessarily reflect those of the European Union or the European Research Council Executive Agency. Neither the European Union nor the granting authority can be held responsible for them. This project was also funded by the Deutsche Forschungsgemeinschaft (DFG, German Research Foundation) under grant numbers EXC 2089/1–390776260 (Germany's Excellence Strategy) and TI 1063/1 (Emmy Noether Program), the Bavarian program Solar Energies Go Hybrid (SolTech) and the Center for NanoScience (CeNS). S.A.M. additionally acknowledges the Lee-Lucas Chair in Physics.

*Author contributions*

J.B., A.T. conceived the idea and planned the research. J.B. and C.H. contributed to the sample fabrication. J.B., D.G. and L.S. performed the optical measurements. J.B., C.H. and T.W. performed the data processing and data analysis. L.S.M., S.A.M and A.T. supervised the project. All authors contributed to the writing of the paper.

**Competing Interests**

Authors declare that they have no competing interests.

# Supporting Information

## Fabrication Optimization of van der Waals Metasurfaces: Inverse Patterning Boosts Resonance Quality Factor

*Jonas Biechteler, Connor Heimig, Thomas Weber, Dmytro Gryb, Luca Sortino, Stefan A. Maier, Leonardo de S. Menezes, Andreas Tittl[*]*

*SI note 1: Lift-off Process Flow*

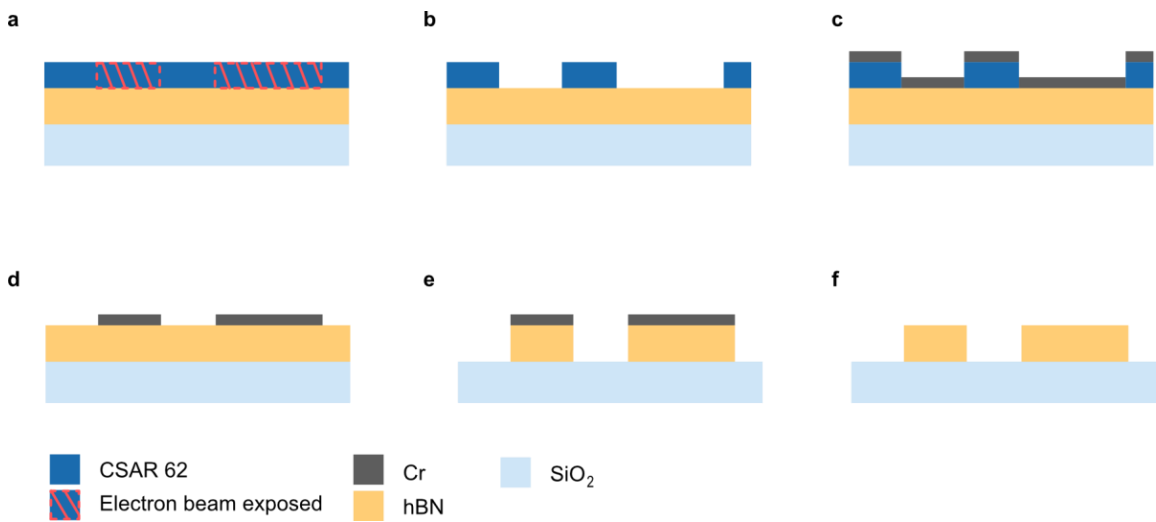

**SI Figure 1. Illustration of common lift-off based fabrication technique. a.** Electron-beam resist (PMMA or CSAR) coating on top of the hexagonal boron nitride (hBN) flake and subsequent electron-beam exposure in later resonator areas. **b.** Development washing away exposed resist and forming resist pillars on later gaps **c.** Metal hard mask coating, here chromium **d.** Wet-chemical lift-off. Resist is dissolved in solvent, removing all chromium film areas without direct contact to substrate or hBN flake. **e.** Reactive-ion etching (RIE) of hBN. **f.** Chromium mask removal again via RIE.

*SI note 2: Influence of Resonator Height*

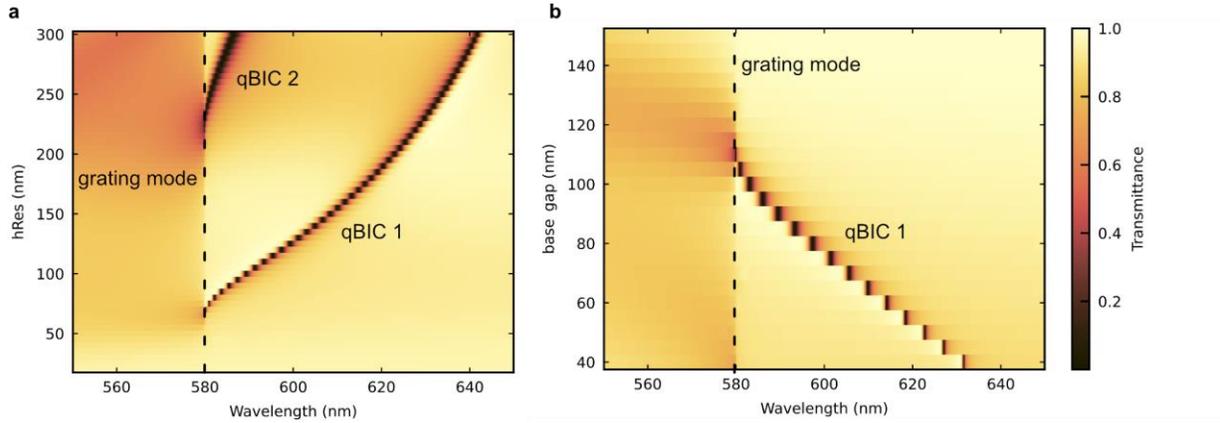

**SI Figure 2. a.** Simulated spectra of metasurfaces with increasing resonator height (hRes) for a xy-periodicity of 400 nm. All other geometric parameters are fixed. Decreasing height causes shrinking in resonator and photonic mode volume and a continuous blue-shift of the resonance. Too little height and mode volume causes loss of all qBIC resonances to the diffraction modes, here marked as grating mode that appears due to the periodicity of the unit cell array forming the metasurfaces. With enough hRes, even higher order qBICs can emerge (qBIC 2). **b.** Simulated spectra of metasurfaces with increasing base gap between all resonators with fixed parameters. Again, increasing gap size causes decreasing mode volume and a blue-shift, until the qBIC resonance is lost to the grating mode.

*SI note 3: TCMT fits*

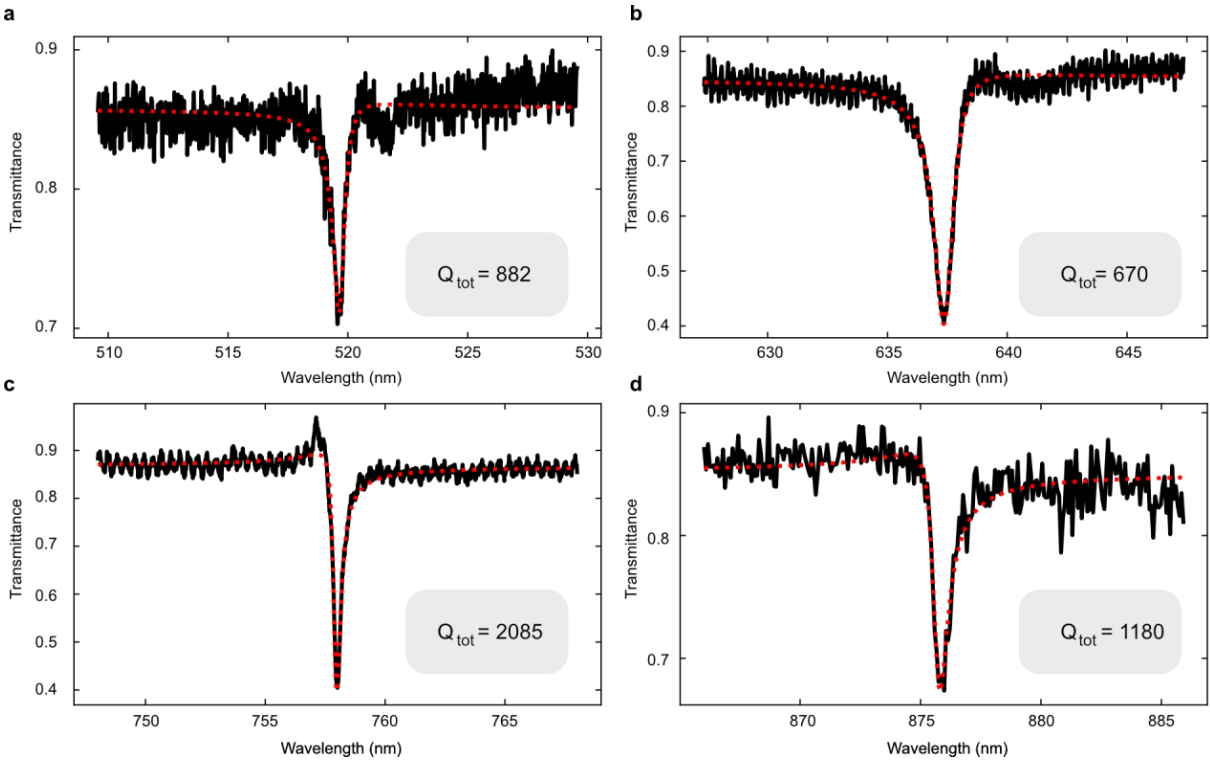

**SI Figure 3.** TCMT fitting of four exemplary qBIC resonance spectra from Figure 3a with total Q factors $Q_{tot}^{-1} = Q_{rad}^{-1} + Q_{int}^{-1}$. Associated S = 0.8 (**a**), S = 1.0 (**b**), S = 1.2 (**c**) and S = 1.4 (**d**).

*SI note 4: LOD calculation*

Calculation of the limit of detection[1] $LOD = 3\sigma * S_B^{-1}$ with σ standard deviation in resonance position was done with continuous measurement of resonance spectra, every 2 seconds for 100 seconds in total. Evaluation of resonance position by centroid wavelength $\lambda_{centroid}$[2], calculated by:

$$\lambda_{centroid}(t) = \frac{\int_{\lambda_s}^{\lambda_s+S} \lambda(\varepsilon_{fit}(\lambda,t) - \varepsilon_{base})d\lambda}{\int_{\lambda_s}^{\lambda_s+S}(\varepsilon_{fit}(\lambda,t) - \varepsilon_{base})d\lambda}$$

Final deviation calculated to be $\sigma = 2.258 * 10^{-3}$ nm.

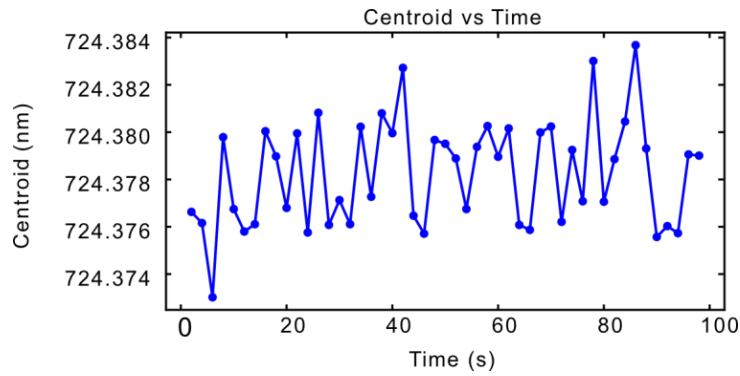

**SI Figure 4.** Centroid wavelength position over time acquired by measurement of resonance spectra over 100 seconds and subsequent TCMT fitting and centroid calculation.

*SI note 5: Maximally chiral hBN metasurfaces in simulation*

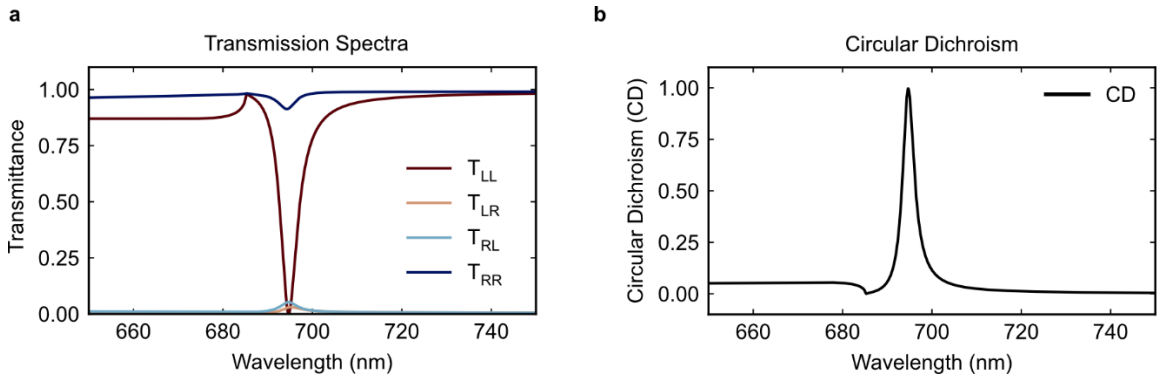

**SI Figure 5.** Simulation of maximally chiral hBN metasurface with two-height geometry. the geometry of the final two-height resonator unit cell (right). The unit cell design has a period of $p_x = 472$ nm and $p_y = 436$ nm, rods tilted by $\theta = 15°$, with lengths of $L_1 = 288$ nm and $L_2 = 288$ nm, widths $W_1 = 121.5$ nm and $W_2 = 170$ nm, and heights $H_1 = 260$ nm and $H_2 = 160$ nm, respectively. **a.** Simulated transmittance spectra of co- and cross-polarized terms. Co-polarized spectra show the emergence of a fully modulated qBIC resonance for only one handedness, cross-polarized terms are insignificant. **b.** Calculated linear circular dichroism (CD) for the simulated spectra. CD reaches up to closely $CD = 100\%$ at the qBIC resonace spectral position.